\documentclass[
 reprint,nofootinbib,
 amsmath,amssymb,
 aps,
]{revtex4-2}

\usepackage{graphicx}
\usepackage{dcolumn}
\usepackage{bm}
\usepackage{xcolor}
\usepackage{diagbox}
\usepackage[colorinlistoftodos]{todonotes}

\usepackage{soul}

\begin{document}

\preprint{APS/123-QED}

\title{Resolution-Independent Machine Learning Heat Flux Closure for ICF Plasmas}

\author{M.~Luo${}^{1}$, A.~R.~Bell${}^{1,2}$, F.~Miniati${}^{3}$, S.~M.~Vinko${}^{1}$ and G.~Gregori${}^{1}$}

\affiliation{%
 ${}^1$ Department of Physics, University of Oxford, Parks Road, Oxford OX1 3PU, UK
}%
\affiliation{%
 ${}^2$ Central Laser Facility, STFC Rutherford Appleton Laboratory, Oxfordshire OX11 0QX, UK
}%
\affiliation{%
 ${}^3$  Mach42, Robert Robinson Avenue, Oxford Science Park, Oxford, OX4 4GP, UK
}%

\begin{abstract}
Accurate modeling of heat flux in inertial confinement fusion plasmas requires closures that remain predictive far from local equilibrium and across disparate spatial and temporal resolutions. We develop a resolution-independent machine-learning heat flux closure trained on particle-in-cell simulations using a Fourier Neural Operator. Two nonlocal electron thermal conduction models are trained and tested. When embedded self-consistently into the electron energy equation, the learned closure faithfully reproduces the temperature evolution and shows good temporal extrapolation and generalization capability. Remarkably, models trained on coarse-resolution data accurately predict heat flux when deployed in substantially finer-resolution implicit, iterative solvers of the energy equation, significantly enhancing the practicality of embedding data-driven closures into partial differential equation solvers. These results establish a data-driven closure that bridges kinetic and fluid descriptions and provides a viable pathway for treating machine learning as an iterative solver within the radiation-hydrodynamic simulations of ICF plasma.
\end{abstract}

\maketitle

Understanding the temperature gradient driven~\cite{Gianluca1} or current driven~\cite{current-driven} heat transport in plasmas is essential for a broad range of applications, from inertial confinement fusion (ICF)~\cite{ICFoverreviw} to astrophysical systems~\cite{silva2018nonlocal, somov2012plasma}. In hot plasmas, the character of temperature gradient driven heat transport is governed by the Knudsen number, which is defined as the ratio of the electron mean free path $\lambda_0$~\cite{huba2023plasma}, to the temperature gradient scale length, $L_T = T_e / \nabla T_e$; here $T_e$ is the electron temperature, and the electron mean free path is $\lambda_0=4 \pi\varepsilon_0^2 m_e^2 v_{\rm {th}}^4/(Z n_ee^4 \ln\Lambda)$~\cite{huba2023plasma}, with $\varepsilon_0$ the vacuum permittivity, $m_e$ the electron mass, $v_{th}=\sqrt{T_e/m_e}$ the electron thermal velocity, $Z$ the ion charge state, $n_e$ the electron number density, $e$ the elementary charge, and $\ln\Lambda$ the Coulomb logarithm. When $\lambda_0/L_T \ll 1$, transport is well described by local theories, and classical models such as the Spitzer-H\"arm (SH) formulation~\cite{SH} yield reliable predictions. As $\lambda_0/L_T$ increases, kinetic and nonlocal effects become increasingly important, leading to the breakdown of local hydrodynamic closures~\cite{nonlocalclosure}. 

Among existing nonlocal theories~\cite{SNB, Chrisment, Michel2023}, the Schurtz-Nicola\"{\i}-Busquet (SNB) model~\cite{SNB} has been widely adopted in radiation-hydrodynamic simulations of ICF and implemented in several ICF codes~\cite{hydra,lilac,fci2,chic}. To further improve agreement with Vlasov-Fokker-Planck (VFP)~\cite{bell} simulations, a number of SNB variants~\cite{snbvariant1,snbvariant2,snbvariant3} have been proposed, primarily through modifications of the effective mean free path within the SNB framework. Despite these efforts, the discrepancies with kinetic simulations persist in strongly nonlocal regimes, and the traditional numerical solver of the SNB model introduces substantial computational overhead when coupled to a radiation-hydrodynamic code, and this extra computational expense motivates the exploration of data-driven approaches~\cite{Lamy2022, mfluoMLP, mfluoCNN, miniati, fiona} as potential alternatives for nonlocal heat-flux closure. However, conventional neural networks, such as multilayer perceptrons and convolutional networks, learn mappings tied to fixed discretizations, which limits their suitability for closures that must be embedded into partial differential equations and remain predictive across varying spatial and temporal resolutions.

In this Letter, we employ a neural operator framework~\cite{neuraloperator} to learn the functional mapping from the electron temperature profile, \(T_e(x)\), to the divergence of the heat flux, \(\partial_x q(x)\), enabling the construction of a resolution-independent closure model. Specifically, we adopt the Fourier Neural Operator (FNO)~\cite{FNOLi}, which exploits spectral representations to reduce the computational complexity to \(\mathcal{O}(n\log n)\) ($n$ is the number of discretized points), while naturally encoding the intrinsically nonlocal character of the heat transport through global interactions in Fourier space. Fully kinetic particle-in-cell (PIC) simulations were performed with the OSIRIS code~\cite{osiris} to generate the data. A comparison between published VFP results, PIC simulations, and the SNB model is provided in the Supplemental Material. Our implementation of the SNB model follows that of Refs.~\cite{snbvariant1,snbvariant3}, ensuring consistency with previous analyses; A total of 180 energy groups is employed, with the uniform energy grid ranging from 0 velocity up to 25 times the highest temperature value, see also the Supplemental Material for further details. The data used to train the model comprises two transport test cases: the relaxation of a \textit{hot spot}~\cite{Batishchev2002}, and the decay of a small-amplitude sinusoidal temperature perturbation, commonly referred to as the \textit{Epperlein–Short} (\textit{ES}) test~\cite{Epperlein}, both considered at constant plasma density. 

Spatial ($x$) and temporal ($t$) coordinates are normalized by the mean free path $\lambda_{0}$ and the corresponding collision time $1/\nu_{0}$, evaluated at a reference density $n_{e0}$ and temperature $T_{e0}$ (in the following, temperature profiles are also normalized by the reference temperature $T_{e0}$, and the heat flux is normalized by the free stream flux $n_{e0}T_{e0}v_{vth0}$, $v_{th0}=\sqrt{T_{e0}/m}$). The Coulomb logarithm is taken to be constant, and the periodic boundary condition is applied. The simulations are performed in a one-dimensional domain of length $L=100$ over a time interval $\tau=30$. The spatial simulation step is $\Delta x=\lambda_{\mathrm{De},0}/2$, where $\lambda_{\mathrm{De},0}$ is the Debye length, and the time step $\Delta t$ is chosen to satisfy the Courant–Friedrichs–Lewy condition, with $\Delta t/\Delta x \approx 0.9$. A fully ionized hydrogen plasma with ion charge state $Z=1$ is assumed. Each cell contains $6.4\times10^{4}$ electron and $6.4\times10^{4}$ ion macroparticles. Simulation data are recorded at spatial and temporal resolutions of $\mathbf{dx}\approx0.06$ and $\mathbf{dt}=0.01$, respectively. 

For the \textit{hot spot} case, the initial temperature profile is prescribed as $T_e(x)=1+\exp[-x^2/(\alpha \ell)^2]$, where $\ell=8.44$ and the parameter $\alpha$ is chosen from the set $\{1,\,1.25,\,1.5,\,1.75,\,2\}$. The Knudsen number reaches values of $\sim0.17$, at which strongly nonlocal behavior of heat transport occurs. In the \textit{ES} case, the initial temperature profile is given by $T_e(x)=1+\delta T_0 \cos(\beta k_0 x)$, where $\delta T_0 = 4\%$, $k_0 = 2\pi/L$, and $\beta \in \{1,\,4,\,7,\,10,\,13,\,16\}$. These choices correspond to $k\lambda_0 = \beta k_0 \lambda_0\in \{0.063,\,0.251,\,0.440,\,0.628,\,0.817,\,1.000\}$; here $k=\beta k_0$ is the perturbation wavenumber. The model learns the operator $\partial_xq=\mathcal{F}_{FNO}(T_e)$ based on PIC simulation data belonging to the time interval $t\in(0,20]$. We expect the trained operator $\mathcal{F}_{\mathrm{FNO}}(T_e)$ to capture the temperature evolution governed by
\begin{equation}
    (3/2)\partial T_e/\partial t+\mathcal{F}_{FNO}(T_e)=0, 
    \label{equ}
\end{equation}
which is solved implicitly using an iterative scheme~\cite{Duccao}. Meanwhile, the saved data in PIC are downsampled using several strategies before being input into the same FNO architecture, to demonstrate learning rather than memorization. These downsampling strategies are shown in Table~\ref{tabel}, where $dx$ and $dt$ denote the spatial and temporal resolutions of the downsampled data. Hence, a total of 15 data groups with different resolutions are independently input into the same FNO architecture, and the performance of each model is evaluated. For convenience, the model trained at a specific resolution, $dx/\mathbf{dx} = n$ and $dt/\mathbf{dt} = m$, is denoted by $\mathcal{F}_{(n,m)}$. The training strategies and the model performance for $T_e \mapsto \partial_xq$ are detailed in the Supplemental Material.

\begin{table}[h]
\caption{Resolutions of the datasets used for training.}
\label{tabel}
\begin{ruledtabular}
\begin{tabular}{c c c c c c}
\diagbox[width=6.0em,height=2.0em]{$dx/\mathbf{dx}$}{$dt/\mathbf{dt}$}
 & 1 & 2 & 3 & 5 & 10 \\
\hline
1 & $(1,1)$ & $(1,2)$ & $(1,3)$ & $(1,5)$ & $(1,10)$ \\
3 & $(3,1)$ & $(3,2)$ & $(3,3)$ & $(3,5)$ & $(3,10)$ \\
6 & $(6,1)$ & $(6,2)$ & $(6,3)$ & $(6,5)$ & $(6,10)$ \\
\end{tabular}
\end{ruledtabular}
\end{table}

By introducing the surrogate model into Eq.~\eqref{equ} via $\mathcal{F}_{\mathrm{FNO}}(T_e)=\mathcal{F}_{(n,m)}(T_e)$, we assess its predictive capability as follows.

\begin{figure}[tbp]
    \centering
    \includegraphics[width=1\linewidth]{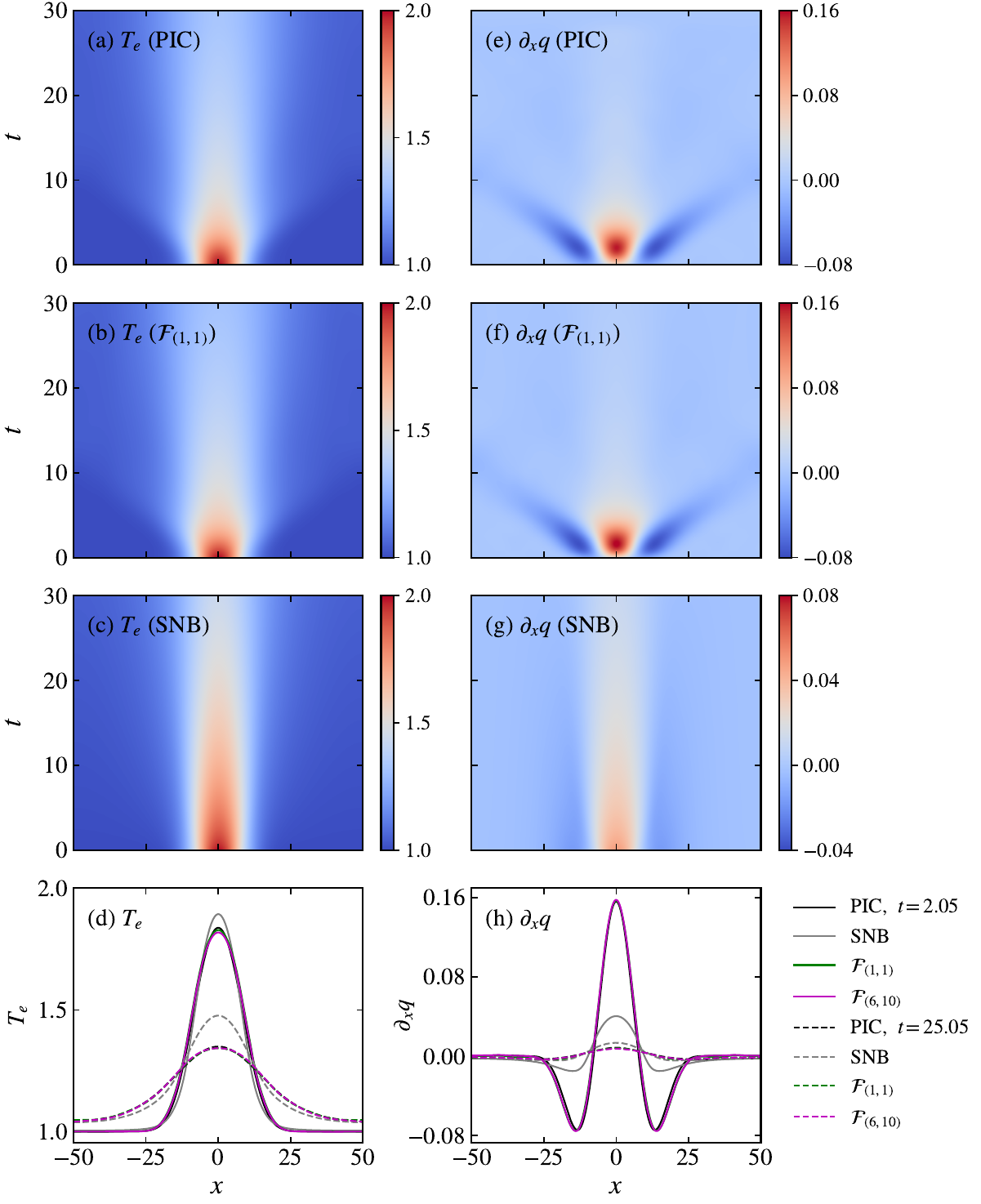}
    \caption{Spatiotemporal evolution of the electron temperature $T_e$ (left column) and the heat flux divergence $\partial_xq$ (right column) for $\alpha=1.125$ of the \textit{hot spot} case. The first, second, and third rows correspond to results from PIC simulations, solutions of Eq.~\eqref{equ} using the FNO model $\mathcal{F}_{(1,1)}$, and solutions obtained using the SNB model, respectively. The bottom row shows line plots of $T_e$ and $\partial_xq$ at $t=2.05$ and $t=25.05$ of the PIC simulations, solutions of Eq.~\eqref{equ} using the FNO model $\mathcal{F}_{(1,1)}$, $\mathcal{F}_{(6,10)}$, and solutions obtained using the SNB model.}
    \label{hotspot1125}
\end{figure}

\textit{Hot spot} test. In addition to the data used to train the model, namely, $\alpha\in\{1,\,1.25,\,1.5,\,1.75,\,2\}$, which define the width of the initial Gaussian temperature profile, the model is also evaluated on unseen intermediate values $\alpha\in\{1.125,\,1.375,\,1.625,\,1.875\}$ to assess generalization. Figure~\ref{hotspot1125} shows the spatiotemporal evolution of the temperature $T_e$ (left column) and the divergence of the heat flux $\partial_xq$ (right column) for $\alpha=1.125$, shown at the resolution $dx=\mathbf{dx}$ and $dt=\mathbf{dt}$ over $t \in (0,30]$. That is, Eq.~\eqref{equ} is solved using $dx=\mathbf{dx}$ and $dt=\mathbf{dt}$. Note the model $\mathcal{F}_{(n,m)}$ is trained using data from $t \in (0,20]$, whereas the interval $t \in (20,30]$ represents the model's predicted future evolution, i.e., temporal extrapolation. The first, second, and third rows correspond to results from PIC simulations, from solving Eq.~\eqref{equ} using the FNO model $\mathcal{F}_{(1,1)}$, and the SNB model, respectively. Please note Eq.~\eqref{equ} is solved implicitly and iteratively, with two iteration cycles performed at each time step for both the $\mathcal{F}_{(1,1)}$ and SNB models. The solution of Eq.~\eqref{equ} using $\mathcal{F}_{(1,1)}$ agrees well with PIC results, whereas the SNB-based solution shows substantial discrepancies. In particular, the SNB model underestimates $\partial_xq$ compared to the PIC, leading to a reduced change in temperature. The bottom row in Fig.~\ref{hotspot1125} shows line plots of $T_e$ and $\partial_xq$ at $t=2.05$ and $t=25.05$ (temporal extrapolation). Besides the solution of Eq.~\eqref{equ} with $\mathcal{F}_{(1,1)}$, the solution of Eq.~\eqref{equ} with $\mathcal{F}_{(6,10)}$ is also plotted. We observe that the solution of Eq.~\eqref{equ} using $\mathcal{F}_{(6,10)}$ still provides accurate predictions despite being trained on coarse-resolution data; especially, these two time indices shown were never used during the training of $\mathcal{F}_{(6,10)}$.

\begin{figure}[b]
    \centering
    \includegraphics[width=1\linewidth]{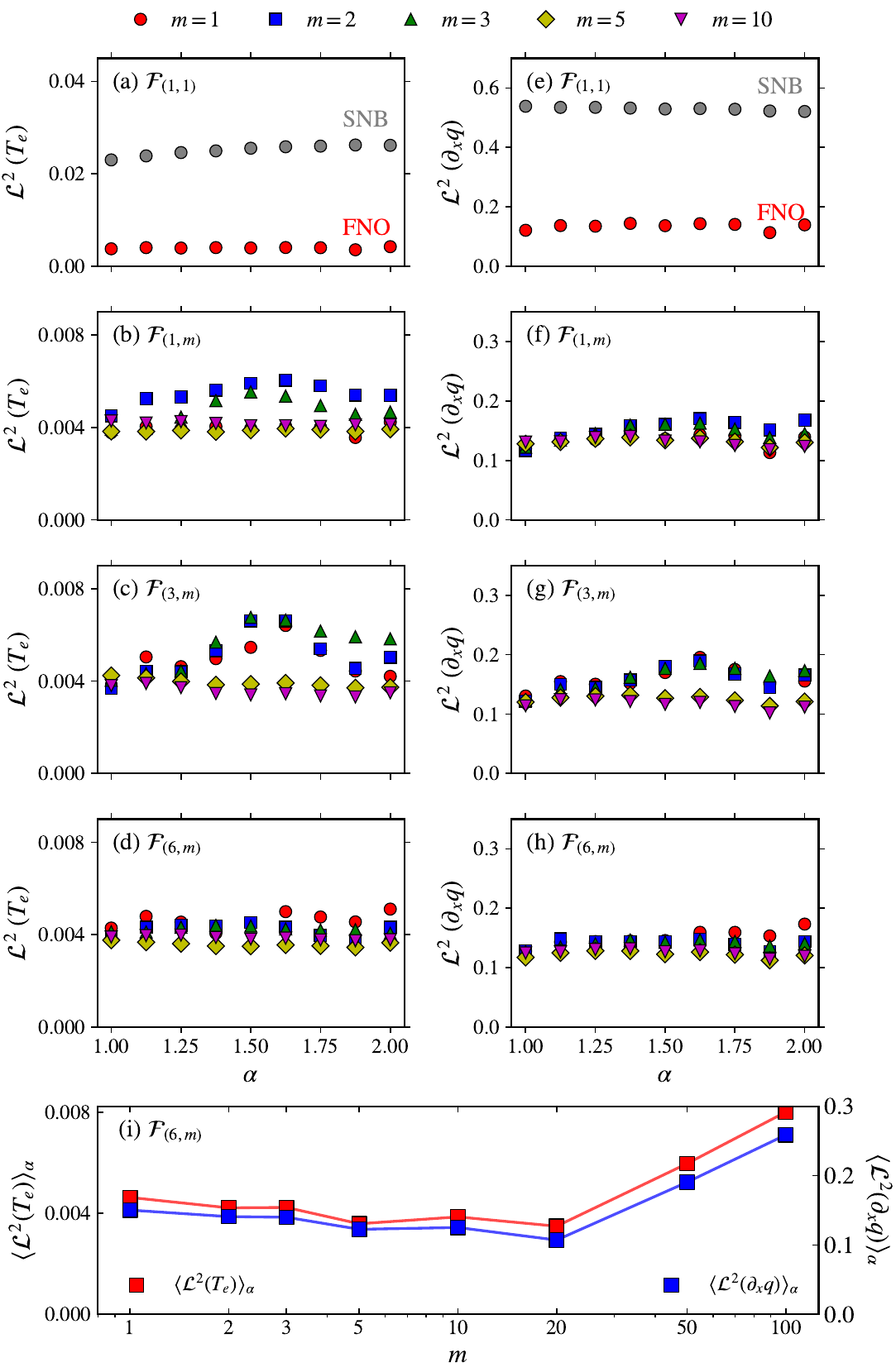}
    \caption{Relative $\mathcal{L}^2$ errors for the \textit{hot-spot} case obtained using different $\mathcal{F}_{(n,m)}$ models. The left and right columns show the relative $\mathcal{L}^2$ errors for $T_e$ and $\partial_xq$, respectively. The top row shows results for the $\mathcal{F}{(1,1)}$-based (red circles) and SNB-based (gray circles) solutions. The second, third, and fourth rows correspond to solutions obtained using $\mathcal{F}_{(1,m)}$, $\mathcal{F}_{(3,m)}$, and $\mathcal{F}_{(6,m)}$, respectively, with different colors indicating $m$=1 (red circles), 2 (blue squares), 3 (green up-triangles), 5 (yellow diamonds), and 10 (magenta down-triangles). The bottom row shows the averaged relative errors $\langle \mathcal{L}^2(T_e) \rangle_{\alpha}$ (left axis) and $\langle \mathcal{L}^2(\partial_x q) \rangle_{\alpha}$ (right axis) of the model $\mathcal{F}_{(6,m)}$ over all $\alpha$ as functions of the temporal resolution ($m=dt/\mathbf{dt}$) in the training data.}
    \label{hotspotall}
\end{figure}

The performance of all $\mathcal{F}_{(n,m)}$ models on \textit{hot spot} case is summarized in Fig.~\ref{hotspotall} and evaluated using the relative $\mathcal{L}^2$ error, defined as $\mathcal{L}^2(\mathcal{E})=(\sum_{x,t}[\mathcal{E}_g-\mathcal{F}_{\mathcal{E}}]^2)^{1/2}/(\sum_{x,t}\mathcal{E}_g^2)^{1/2}$. Here, $\mathcal{E}$ denotes either $T_e$ or $\partial_xq$, the subscript $g$ indicates the ground truth from PIC results, and $\mathcal{F}_{\mathcal{E}}$ represents the solution of Eq.~\eqref{equ} obtained using the corresponding $\mathcal{F}_{(n,m)}$. $\mathcal{L}^2(T_e)$ and $\mathcal{L}^2(\partial_xq)$ are shown in the left and right columns of Fig.~\ref{hotspotall}, respectively. The top row presents the $\mathcal{L}^2(\mathcal{E})$ of the $\mathcal{F}{(1,1)}$-based (red circles) and SNB-based (gray circles) solutions of Eq.~\eqref{equ}, showing the superior accuracy of $\mathcal{F}{(1,1)}$-based solution, as already observed in Fig.~\ref{hotspot1125}. The $\mathcal{L}^2(\mathcal{E})$ of the solutions obtained using $\mathcal{F}_{(1,m)}$, $\mathcal{F}_{(3,m)}$, and $\mathcal{F}_{(6,m)}$ are shown in the second, third, and fourth rows, respectively, with different colors indicating $m$=1 (red circles), 2 (blue squares), 3 (green up-triangles), 5 (yellow diamonds), and 10 (magenta down-triangles). The $\mathcal{L}^2(\mathcal{E})$ remains stable across $\alpha$ and is largely insensitive to the choice of $\mathcal{F}_{(n,m)}$. Notably, models trained on coarse-resolution data (e.g., $\mathcal{F}_{(6,10)}$) still achieve high accuracy when deployed within the fine resolution solver of Eq.~\eqref{equ}. Fig.~\ref{hotspotall}(i) shows the averaged relative errors $\langle \mathcal{L}^2(T_e) \rangle_{\alpha}$ (left axis) and $\langle \mathcal{L}^2(\partial_x q) \rangle_{\alpha}$ (right axis) of the model $\mathcal{F}_{(6,m)}$ over all $\alpha$ as functions of the temporal resolution in the training data. We find even for a coarser resolution of $dt/\mathbf{dt}=20$, the model $\mathcal{F}_{(6,20)}$ retains good predictive performance. In addition, solving Eq.~\eqref{equ} using $\mathcal{F}_{(6,10)}$ with larger time steps, $dt=10\mathbf{dt}$ and $dt=100\mathbf{dt}$, still yields good agreement with the PIC results (shown in the Supplemental Material), although more iterations are required to advance each time step.

\begin{figure}[t]
    \centering
    \includegraphics[width=0.95\linewidth]{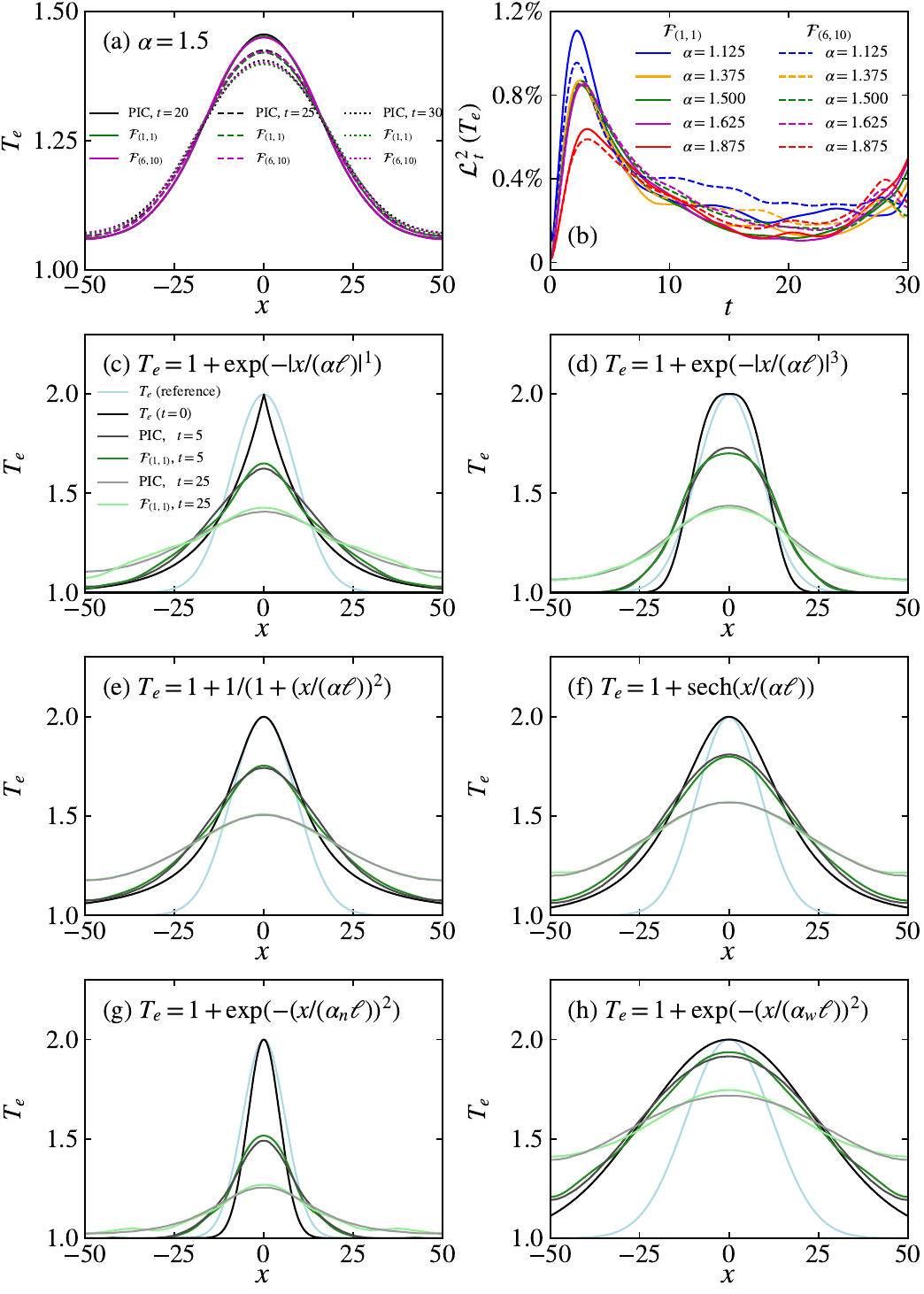}
    \caption{Assessment of temporal extrapolation and generalization of the FNO model. (a) Temperature evolution obtained from PIC simulations (black) and from solutions of Eq.~\eqref{equ} using the FNO models $\mathcal{F}{(1,1)}$ (green) and $\mathcal{F}{(6,10)}$ (magenta) for an initial Gaussian profile with $\alpha=1.5$. Profiles at $t=20$ (solid), $t=25$ (dashed), and $t=30$ (dotted) are shown. (b) Temporal relative $\mathcal{L}^2$ error of the temperature, $\mathcal{L}^2_t(T_e)$, for $\alpha\in\{1.125,1.375,1.5,1.625,1.875\}$ using $\mathcal{F}{(1,1)}$ (solid, trained with the highest-resolution data) and $\mathcal{F}{(6,10)}$ (dashed, trained with the lowest-resolution data). (c)–(f) Model performance for different initial temperature profiles (black): sub-Gaussian $T_e=1+\exp[-|x/(\alpha\ell)|^1]$, super-Gaussian $T_e=1+\exp[-|x/(\alpha\ell)|^3]$, Lorentzian-type $T_e=1+1/[1+(x/(\alpha\ell))^2]$, and $T_e=1+\mathrm{sech}(x/(\alpha\ell))$, respectively, with $\alpha=1.5$. The reference Gaussian profile is shown in light blue. (g),(h) Performance on Gaussian profiles outside the training range with narrower ($\alpha_n=0.75$) and broader ($\alpha_w=4$) widths. Light-blue curves indicate the narrowest ($\alpha=1$) and broadest ($\alpha=2$) Gaussian profiles included in the training set. Temperature profiles are shown at $t=5$ (strong nonlocality) and $t=25$ (less nonlocality).}
    \label{generalize}
\end{figure}

To further assess the model’s temporal extrapolation capability, Fig.~\ref{generalize}(a) shows the temporal evolution of the temperature obtained from PIC simulations (black) and from solutions of Eq.~\eqref{equ} using the FNO models $\mathcal{F}{(1,1)}$ (green) and $\mathcal{F}{(6,10)}$ (magenta). Results are shown at $t=20$ (solid), $t=25$ (dashed), and $t=30$ (dotted) for an initial Gaussian temporal profile with $\alpha=1.5$. Excellent agreement between the PIC results and the FNO predictions is observed. Meanwhile. the temporal relative $\mathcal{L}^2$ error of the temperature, defined as $\mathcal{L}^2_t(T_e)=(\sum_{x}[T_e(t)_g-\mathcal{F}_{T_e(t)}]^2)^{1/2}/(\sum_{x}T_e(t)_g^2)^{1/2}$, is shown in Fig.~\ref{generalize}(b) with $\alpha\in\{1.125,\,1.375,\,1.5,\,1.625,\,1.875\}$, and model $\mathcal{F}_{(1,1)}$ (solid, trained using the highest resolution data) and $\mathcal{F}_{(6,10)}$ (dashed, trained using the lowest resolution data) are examined. Over nearly the entire time domain, the relative error $\mathcal{L}^2_t(T_e)$ remains below $1\%$, including during the temporal extrapolation interval $t \in (20,30]$, demonstrating the high predictive accuracy of the learned model. Furthermore, the error curves from $\mathcal{F}{(1,1)}$ (solid) and $\mathcal{F}{(6,10)}$ (dashed) for the same $\alpha$ show comparable behavior, except in the far-tail region, further indicating the resolution independence of the training and the performance.

Figures~\ref{generalize}(c)–(f) show the performance of the model $\mathcal{F}_{(1,1)}$ for different initial temperature profiles (black curves): a sub-Gaussian profile $T_e = 1+\exp[-|x/(\alpha \ell)|^1]$ in (c), a super-Gaussian profile $T_e = 1+\exp[-|x/(\alpha \ell)|^3]$ in (d), a Lorentzian-type profile $T_e = 1 + 1/[1+(x/(\alpha \ell))^2]$ in (e), and $T_e = 1 + \mathrm{sech}(x/(\alpha \ell))$ in (f). Here $\alpha = 1.5$, and the reference Gaussian profile is also shown in light blue. Figures~\ref{generalize}(g) and (h) further test the model on Gaussian profiles with narrower ($\alpha_n = 0.75$) and broader ($\alpha_w = 4$) widths. The light-blue curves in (g) and (h) correspond to the narrowest ($\alpha = 1$) and broadest ($\alpha = 2$) Gaussian profiles included in the training data. The temperature profiles at two time instances, $t=5$ (strong nonlocality) and $t=25$ (less nonlocality), are shown. The solutions of Eq.~\eqref{equ} obtained with $\mathcal{F}_{(1,1)}$ show good agreement with PIC simulations, demonstrating the model’s generalization capability.

\begin{figure}[b]
    \centering
    \includegraphics[width=1\linewidth]{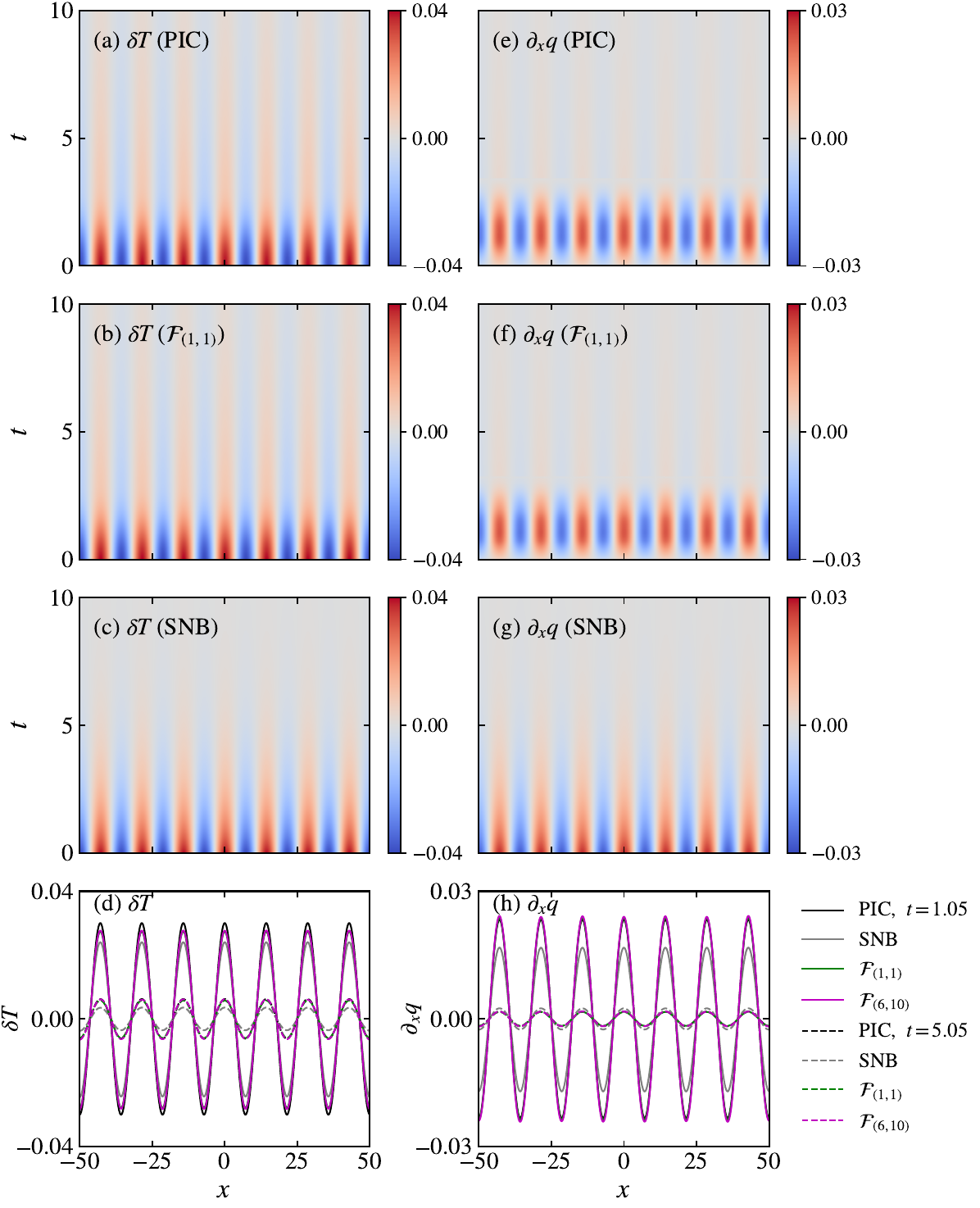}
    \caption{Spatiotemporal evolution of the temperature $T_e$ and the corresponding heat flux divergence $\partial_xq$ for the \textit{ES} case with $\beta=7$ or $k\lambda_0 \approx 0.44$. The first, second, and third rows show results from PIC simulations, solutions of Eq.~\eqref{equ} obtained using the FNO model $\mathcal{F}{(1,1)}$, and solutions obtained using the SNB model, respectively. The bottom row shows line plots of $T_e$ and $\partial_xq$ at $t=1.05$ and $t=5.05$ of the PIC simulations, solutions of Eq.~\eqref{equ} using the FNO model $\mathcal{F}_{(1,1)}$, $\mathcal{F}_{(6,10)}$, and solutions obtained using the SNB model.}
    \label{es7}
\end{figure}

\textit{ES} test. We next examine the performance of all $\mathcal{F}_{(n,m)}$ models for the \textit{ES} case. The temperature decay (left column: $\delta T$) and corresponding evolution of the heat flux divergence (right column: $\partial_xq$) for the case $\beta$=7, leading to $k\lambda_0\approx$0.44, are shown in Fig.~\ref{es7}. The first, second, and third rows correspond to results from PIC simulations, from solving Eq.~\eqref{equ} using the FNO model $\mathcal{F}_{(1,1)}$, and the SNB model, respectively. Again, the PIC ground-truth is shown at the reference resolution $dx=\mathbf{dx}$ and $dt=\mathbf{dt}$, and Eq.~\eqref{equ} is solved at the same resolution. The solution of Eq.~\eqref{equ} obtained using $\mathcal{F}_{(1,1)}$ exhibits high accuracy, whereas the solution obtained with the SNB model shows substantial deviations from the PIC results. The bottom row of Fig.~\ref{es7} shows line plots of $\delta T$ and $\partial_xq$ at $t=1.05$ and $t=5.05$. We compare the solutions of Eq.~\eqref{equ} obtained with \(\mathcal{F}_{(6,10)}\) and \(\mathcal{F}_{(1,1)}\). Despite being trained on coarser data, \(\mathcal{F}_{(6,10)}\) produces solutions that remain quantitatively accurate. We again emphasize that these time instances were not included in the training of $\mathcal{F}_{(6,10)}$.

\begin{figure}[b]
    \centering
    \includegraphics[width=0.95\linewidth]{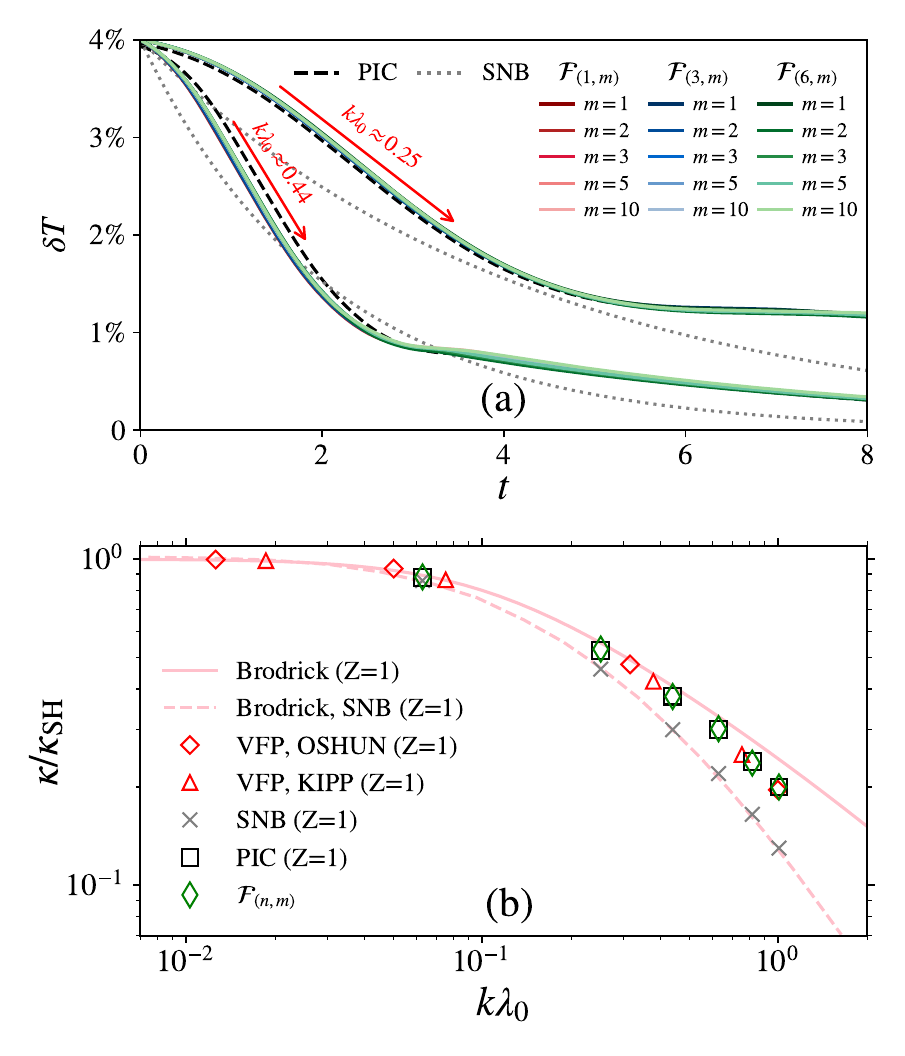}
    \caption{(a) Temporal decay of the temperature perturbation $\delta T(t)$ for the \textit{ES} cases with $\beta=4$ and $7$, corresponding to $k\lambda_0 \approx 0.25$ and $0.44$, respectively. Black dashed curves show PIC results, gray dotted curves show solutions of Eq.~\eqref{equ} obtained using the SNB model, and colored curves correspond to solutions obtained using the $\mathcal{F}{(n,m)}$ models trained at different resolutions (red for $\mathcal{F}_{(1,m)}$, blue for $\mathcal{F}_{(3,m)}$, and green for $\mathcal{F}_{(6,m)}$). (b) Effective thermal conductivity ratio $\kappa/\kappa_{\rm SH}$ as a function of $k\lambda_0$. The red diamonds and red triangles denote published results from the VFP codes OSHUN~\cite{snbvariant2} and KIPP~\cite{snbvariant3}, respectively. The solid and dashed pink lines correspond to the fitting function in Eq.~(25) and the SNB model reported in Ref.~\cite{snbvariant3}. The gray crosses indicate results obtained by solving Eq.~\eqref{equ} using the SNB model applied here. And the black squares and green diamonds show $\kappa_{\rm PIC}/\kappa_{\rm SH}$ and $\kappa_{\mathcal{F}_{(n,m)}}/\kappa_{\rm SH}$, respectively, extracted from the early time decay rates of the temperatures in (a).}
    \label{essummary}
\end{figure}

A key quantity in the \textit{ES} case is the decay rate $\gamma$ of the temperature perturbation $\delta T(t)$, defined through $\delta T(t)\propto\exp{(-\gamma t)}$, and the decay rate can be given as $\gamma=2k^2\kappa/3n_{e0}$, where $k$ is the perturbation wavenumber and $\kappa$ is the thermal conductivity. Consequently, the effective thermal conductivity normalized to the SH  thermal conductivity can be inferred from the ratio $\kappa/\kappa_{\rm SH}=\gamma/\gamma_{\rm SH}$. When the SH heat flux closure is employed, the corresponding decay rate satisfies $\gamma_{\rm SH}/\nu_0=(2/3)(k\lambda_0)^2(128/\sqrt{2\pi})(Z+0.24)/(Z+4.2)$.  Figure~\ref{essummary}(a) shows two representative examples of temperature decay corresponding to $\beta=4$ and $7$, i.e., $k\lambda_0 \approx 0.25$ and $0.44$, respectively. The black dashed curves indicate the PIC results, while the gray dotted curves correspond to solutions of Eq.~\eqref{equ} obtained using the SNB model. In both cases, the temperature continues to decay, as the heat flux remains driven by a persistent temperature gradient throughout the time window shown. However, the PIC results exhibit a slower decay at very early times, a more rapid decay at intermediate times, and a slower late-time evolution compared with the SNB model. Regarding the $\mathcal{F}_{(n,m)}$-based solutions of Eq.~\eqref{equ}, indicated by different colors (red for $\mathcal{F}_{(1,m)}$, blue for $\mathcal{F}_{(3,m)}$, and green for $\mathcal{F}_{(6,m)}$), all consistently reproduce the PIC temperature decay, irrespective of the resolution of the training data.

Figure~\ref{essummary}(b) shows the conductivity ratio $\kappa/\kappa_{\rm SH}$ obtained from different models. The red diamonds and triangles denote published results from the VFP codes OSHUN~\cite{snbvariant2} and KIPP~\cite{snbvariant3}, respectively, while the solid and dashed pink lines correspond to the fitting function in Eq.~(25) and the SNB model reported in Ref.~\cite{snbvariant3}. The solution of Eq.~\eqref{equ} obtained using the SNB model applied here, shown by gray crosses, coincides with the dashed pink line. Importantly, both $\kappa_{\rm PIC}/\kappa_{\rm SH}$ (black squares) and $\kappa_{\mathcal{F}{(n,m)}}/\kappa_{\rm SH}$ (green diamonds), extracted from the intermediate-time decay rates shown in Fig.~\ref{essummary}(a), exhibit trends consistent with the VFP results.

In summary, this Letter demonstrates a pathway to replace the conventional SNB heat flux closure with a machine-learning operator trained directly on first-principles simulations. Two representative test cases, the \textit{hot spot} and \textit{ES} problems, are investigated. Although the \textit{hot spot} and \textit{ES} problems differ significantly in amplitude and modulation, the learned operator captures the essential kinetic features of nonlocal heat transport of both problems while enabling fast and stable iterative solutions (for one test case, including both the \textit{hot spot} and \textit{ES} configurations, solving Eq.~\eqref{equ} implicitly and iteratively with the SNB model, requires approximately 800 minutes on a single CPU when integrated to $t=30$ using a time step $dt=\mathbf{dt}$. In contrast, replacing the SNB closure with the learned operator $\mathcal{F}_{(n,m)}$ reduces the time to about 20 minutes under identical conditions, corresponding to a speedup of nearly a factor of 40). At the same time, the learned model shows good temporal extrapolation and generalization capability. Importantly, models trained on coarse-resolution data can be seamlessly deployed within electron energy solvers at arbitrary resolution, including finer grids, significantly enhancing the practicality of embedding data-driven closures into partial differential equation solvers. Meanwhile, training on coarse data further improves the model training efficiency and reduces data requirements without compromising the accuracy of electron temperature evolution. In practice, solving Eq.~\eqref{equ} at the time resolution $dt=\textbf{dt}$ requires $6000$ iterations, corresponding to $3000$ time steps with two iterations per step, while only $2.3\%$ ($1.1\%$) of these profiles are used to train the model $\mathcal{F}_{(n,10)}$ ($\mathcal{F}_{(n,20)}$). \textcolor{black}{However, the model’s predictive capability remains limited when the initial conditions deviate substantially from those represented in the training set. For example, a model trained on \textit{hot spot} cannot reliably predict the \textit{ES} case, as such extrapolation leads to large generalization errors for conditions outside the training distribution. More broadly, robust generalization across widely separated initial conditions remains challenging. Nevertheless, our study represents an important step toward replacing complex closures in hydrodynamic simulations and highlights the potential of treating machine learning as an iterative solver rather than a black-box model.}

In future work, \textcolor{black}{to address the robust generalization challenge,} we aim to learn the underlying first-order spherical harmonic components and to reconstruct the nonlocal heat flux indirectly through the corresponding corrections to the SH flux \textcolor{black}{within the framework of multiple energy groups}. In addition, we will further investigate how externally applied and self-generated magnetic fields influence the internal representations of machine-learning models within the magnetized SNB framework~\cite{msnb}. Both directions critically rely on access to high-fidelity VFP data.

\textit{Acknowledgments}---The authors thank the computing resources provided by the STFC Scientific Computing Department’s SCARF cluster. This work was supported by EPSRC and First Light Fusion under the AMPLIFI prosperity partnership, Grant No. EP/X025373/1.

\textit{Data availability}---The data are not publicly available. The data are available from the authors upon reasonable request.

\bibliography{reference}

\end{document}